\documentclass[twocolumn,reprint,aps,prd,nofootinbib,floatfix,superscriptaddress]{revtex4-1}
\usepackage[utf8]{inputenc}
\usepackage[utf8]{inputenc}
\usepackage{amsmath}
\usepackage{amsfonts}
\usepackage{amssymb}
\usepackage[left=2cm,right=2cm,top=2cm,bottom=2cm]{geometry}
\usepackage{braket}
\usepackage{dsfont}
\usepackage{hyperref}

\usepackage{graphicx}
\usepackage{tikz}
\usetikzlibrary{arrows,decorations.pathmorphing,backgrounds,positioning,fit,petri,shapes}
\usepackage{lipsum}

\newcommand\blfootnote[1]{%
  \begingroup
  \renewcommand\thefootnote{}\footnote{#1}%
  \addtocounter{footnote}{-1}%
  \endgroup
}

\usepackage{multirow}
\begin{document}

%\title{Sorting Redshift Space Distortion Scales in Projection}
\title{The Covariance of Photometric and Spectroscopic Two-Point Statistics: Implications for Cosmological Parameter Inference}

\begin{abstract}
To combine information from measurements of the redshift-space power spectrum from spectroscopic data with angular weak lensing, galaxy clustering and galaxy-galaxy lensing power spectra from photometric surveys (i.e. the $3 \times 2$ point statistics), we must account for the covariance between the two probes. Currently any covariance between the two types of measurements is neglected as existing photometric and spectroscopic surveys largely probe different cosmological volumes. This will cease to be the case as data arrives from Stage-IV surveys. In this paper we derive an analytic expression for the covariance between photometric 2D angular power spectra and the 3D redshift-space power spectrum for Gaussian fields under the plane-parallel approximation. We find that the two probes are covariant on large radial scales, but because the information content of these modes is extremely low due to sample variance, we forecast that it is safe to neglect this covariance when performing cosmological parameter inference.

\end{abstract}

\author{Peter L.~Taylor}
\email{peter.taylor@jpl.nasa.gov}
\affiliation{Jet Propulsion Laboratory, California Institute of Technology, 4800 Oak Grove Drive, Pasadena, CA 91109, USA}
\blfootnote{© 2022. California Institute of Technology. Government sponsorship acknowledged.} 
\author{Katarina Markovi\v{c}}
\affiliation{Jet Propulsion Laboratory, California Institute of Technology, 4800 Oak Grove Drive, Pasadena, CA 91109, USA}

%\date{23 November 2018}
\maketitle

\section{Introduction}
Spectroscopic galaxy clustering measurements~\cite{Reid:2012sw, Reid:2012sw, Macaulay:2013swa, Beutler:2013yhm, Gil-Marin:2015sqa, Simpson:2015yfa, blake2011wigglez, Macaulay:2013swa, Alam:2020sor, Guzzo:2008ac} and the combination of photometric weak lensing, galaxy clustering and galaxy-galaxy lensing~\cite{Asgari:2020wuj, Hikage:2018qbn, Troxel:2017xyo, DES:2021wwk, Heymans:2020gsg} have placed some of the tightest cosmological constraints to date. These probes are primary science targets for many of this decade's largest `Stage-IV' surveys including: Euclid\footnote{\url{http://euclid-ec.org}}~\cite{Blanchard:2019oqi,Laureijs:2011gra}, the Nancy Grace Roman Space Telescope\footnote{\url{https://www.nasa.gov/roman}}~\cite{spergel2015wide}, the Dark Energy Spectroscopic Instrument (DESI)\footnote{\url{https://www.desi.lbl.gov/}}~\cite{Aghamousa:2016zmz} and the Vera Rubin Observatory's Legacy Survey of Space and Time (LSST)\footnote{\url{https://www.lsst.org/}}\cite{LSSTDarkEnergyScience:2018jkl}. 
\par To extract information from spectroscopic measurements of galaxy clustering in 3D at the two-point level, one typically uses a statistic derived from the anisotropic power spectrum, $P(k_\parallel, k_\perp)$ \footnote{The compressed Legendre multipoles $P^i_\ell (k)$ or two-point correlation function  $\xi(s_\perp, s_\parallel)$ are common choices.}. Here we distinguish between radial and perpendicular modes respectively written, $k_\parallel$, and $k_\perp$, to account for anisotropy induced by redshift-space distortions~\cite{kaiser1987clustering} and the Alcock-Paczynski effect~\cite{alcock1979evolution}. 
\par Meanwhile to extract information from photometric data sets we use projected tomographic angular power spectra in 2D, $\{ C_{\rm LL}^{ij}(\ell), C_{\rm GL}^{ij}(\ell), C_{\rm GG}^{ij}(\ell) \}$\footnote{In configuration space, one could alternatively use the projected tomographic angular correlation functions $\{ \xi_{\pm}^{ij}(\theta), \xi_{t}^{ij}(\theta), \xi_{ij}(\theta) \}$.}, where ${\rm LL}, {\rm GL}, {\rm GG}$ denote the cosmic shear, galaxy-galaxy lensing and photometric galaxy clustering signals respectively, while $\{i,j \}$ label the tomographic redshift bins. 
\par To place the tightest cosmological constraints from these next generation experiments, it is imperative to combine the cosmological information from all `Stage-IV' experiments.  Currently there is little sky overlap between photometric and spectroscopic surveys and hence minimal covariance between the two probes. Therefor it is permissible to combine the photometric and spectroscopic parameter constraints by `multiplying the likelihoods' as in~\cite{Heymans:2020gsg, DES:2021wwk}. 
\par However we can anticipate large overlaps between Stage-IV photometric and spectroscopic surveys including Euclid/DESI in the North, Euclid/DESI/LSST around the equator, and Euclid/Roman/LSST in the South. Because photometric and spectroscopic surveys will survey much of the same cosmological volume out to $z \approx 2$, we must correctly account for the covariance between the two measurements to avoid double counting modes. 
\par To combine the two measurements, it may seem natural to extract information from the spectroscopic data set using tomographic angular power spectra~\cite{Gebhardt:2020imr, Jalilvand:2019brk,Joudaki:2017zdt, Loureiro:2018qva}
by dividing the spectroscopic survey window into narrow radial bins to extract radial information and compute the resulting tomographic power spectra as suggested in~\cite{Camera:2018jys}. However, as shown in~\cite{Taylor:2021bhg} this mixes independent radial modes, leading to a large loss of information. 
\par An alternative strategy is to apply a radially harmonic weighting ~\cite{Taylor:2021bhg} (or spherical-Bessel weighting~\cite{Gebhardt:2021mds, Heavens:1994iq, Passaglia:2017lnq}). While promising, this approach has not yet been applied to data and the spherical-Bessel approach is extremely computational expensive deep into the nonlinear regime. Furthermore much more infrastructure exists for measurements of $P(k_\parallel, k_\perp)$ or its derived statistics e.g. the related Legendre multipoles, $P^i_\ell (k)$.
\par Hence we would like to use the anisotropic power spectrum (or Legendre multipoles) to extract information from the spectroscopic survey, and angular power to extract information from the photometric survey. This raises two questions:
\begin{itemize}
    \item {What is the covariance between the 3D anisotropic power spectrum, $P(k_\parallel, k_\perp)$, and the 2D photometric angular power spectra,  $\{ C_{\rm LL}^{ij}(\ell), C_{\rm GL}^{ij}(\ell), C_{\rm GG}^{ij}(\ell) \}$?}
    \item {What impact does accounting for this covariance have on the resulting cosmological parameter constraints?}
\end{itemize}
The objective of this paper is to answer these questions. 
\par To do this, we start by deriving and computing an analytic expression for the covariance between the anisotropic power spectrum and the photometric power spectra for Gaussian fields under the plane-parallel approximation in Sect~\ref{sec:form}. Then in Sect~\ref{sec:results}, we perform a Fisher analysis to compare parameter constraints found with and without the inclusion of the covariance between the photometric and spectroscopic measurements. 
\par The main findings of this paper is that the covariance between $P(k_\parallel, k_\perp)$  and  $\{ C_{\rm LL}^{ij}(\ell), C_{\rm GL}^{ij}(\ell), C_{\rm GG}^{ij}(\ell) \}$ is negligible and it is safe to ignore this covariance during parameter inference. The intuition behind this result is discussed in Sect~\ref{sec:results}.

\section{Formalism} \label{sec:form}
In all that follows we take $\Omega_m = 0.315$, $\Omega_b = 0.04$, $h_0 = 0.67$, $n_s = 0.96$ and $\sigma _8 = 0.8$ as the fiducial cosmology. The matter power spectrum and cosmological distances are computed using {\tt pyCAMB}~\cite{Lewis:1999bs} and we assume the {\tt Halofit} model of~\cite{Takahashi:2012em} to generate the nonlinear power.

\subsection{The Anisotropic Power Spectrum}
We decompose the redshift-space distortion (RSD) spectrum into isotropic and anisotropic parts~\cite{Gebhardt:2020imr}
\begin{equation} \label{eq:pkmu}
P(k_\parallel, k_\perp) =\widetilde{A}^2_{\mathrm{RSD}}(k_\parallel, k_\perp) P(k),
\end{equation}
where $P(k)$ is the matter power spectrum and $\widetilde{A}$ is the RSD operator which accounts for galaxy bias and redshift-space corrections. Here and throughout the remainder of the text, the tilde will indicate quantities and operators in Fourier space. As in~\cite{Gebhardt:2020imr}, we decompose the operator
\begin{equation}
\widetilde{A}_{\mathrm{RSD}}(k_\parallel, k_\perp)= b_g \left(1+\beta \mu^{2}\right) \widetilde{A}_{\mathrm{nl}}(k_\parallel, k_\perp),
\end{equation}
where $b_g \left(1+\beta \mu^{2}\right)$ is the Kaiser term~\cite{kaiser1987clustering} acting on linear scales, $\mu = k_\parallel /k$, $b_g$ is the linear galaxy bias, $\beta = f/b_g$ and $\widetilde{A}_{\mathrm{nl}}$ is the nonlinear redshift-space distortion operator.
In this paper we use a phenomenological Gaussian FoG model to account for the nonlinear redshift-space distortions in the spectroscopic model. In this model~\cite{Hamilton:1997zq}
\begin{equation} \label{eqn:gauss FoG}
\widetilde{A}_{\rm Gauss} (k_\parallel, k_\perp)=  \exp \left[-\frac{1}{2} \sigma_{v}^{2} k_\parallel ^ 2 \right].
\end{equation}
This phenomenological model is not accurate enough for data analysis, but is sufficient for the forecasting work in this paper.
\par Unless explicitly stated otherwise, we take the galaxy bias of the photometric sample $b^{G}_g = 1.5$, the galaxy bias of the spectroscopic sample $b^{s}_g = 1.5$ and $\sigma_{v} = 5 \ h^{-1} {\rm Mpc} $  for the spectroscopic sample closely matching the value chosen in~\cite{Taylor:2021bhg} at $z = 0.675$. For the photometric sample, we ignore the impact of the FoG as the effective velocity dispersion of the photometric redshift is significantly larger rendering the photometric sample insensitive to the FoG.

\subsection{Photometric Angular Power Spectra in the Plane-Parallel Approximation}
In this subsection we will closely follow the derivation in~\cite{Gebhardt:2020imr} which assumes the plane-parallel approximation so that we can relate spherical-harmonic modes, $\ell$, to perpendicular scales, $k_\perp$ following $\ell + 1/2= k_\perp r_0$, where $r_0$ is the effective co-moving distance to the field from Earth. This approximation is valid at the percent-level for $\ell \gtrsim 10$ assuming the radial separation between sources is small (see e.g~\cite{Gebhardt:2020imr, Jalilvand:2019brk, Matthewson:2021rmb}). 
\par We write the observed projected field, $f$, as an integral of some underlying field, $\mathcal{U}^f(\boldsymbol{x})$, along the line-of-sight so that,
\begin{equation} \label{eq:eq1}
    f (\boldsymbol{x_\perp}) = \int_0^{r_{\rm max}} {\rm d}r \ Q^f(r) \mathcal{U}^f (\boldsymbol{x}),
\end{equation}
where $r$ is the co-moving distance, $r_{\rm max}$ is the maximum co-moving distance in the survey and $\boldsymbol{x}$ is the co-moving coordinate so that $r = \boldsymbol{x_\parallel}$ under the plane-parallel approximation. In Fourier space, it is convenient to write the underlying field, $ \widetilde{\mathcal{U}}^f(\boldsymbol{k})$, as a product of a pre-factor, $\mathcal{P}^f$, and the Fourier space matter density contrast, $ \delta(\boldsymbol{k})$, so that
\begin{equation}
   \widetilde{\mathcal{U}}^f(\boldsymbol{k}) = \mathcal{P}^f \delta(\boldsymbol{k}),
\end{equation}
where we take the pre-factors, $\mathcal{P}^f$, to be
    \begin{equation}
     P^{f}=\left\{
  \begin{array}{@{}ll@{}}
    b^G_g  \left(1+\beta \mu^{2} \right), & \text{for} \ f = G \\
    b^G_s \widetilde A_{\rm Gauss} \left(1+\beta \mu^{2}\right), & \text{for} \ f = s \\
    1, & \text{for} \ f = L, \\

  \end{array}\right.
\end{equation}
and $s$, ${ G}$ and ${ L}$ denote spectroscopic clustering, photometric clustering and lensing respectively. Meanwhile the kernel, $Q^f(r)$, is given by
\begin{equation}
     Q^f(r)=\left\{
  \begin{array}{@{}ll@{}}
    n_{\rm G}(r), & \text{for}\ \ f= G\\
    q(r), & \text{for} \ f= L,
  \end{array}\right.
\end{equation}
where $n_{\rm G}(r)$ is the radial distribution function for a tomographic galaxy clustering bin and $q(r)$ is the lensing efficiency kernel defined as
\begin{equation}
   q(r) =  \frac{3}{2} \Omega_m \left( \frac{H_0}{c}\right) ^ 2 \frac{r}{a} \int_r^{r_{\rm max}} {\rm d } r' n_{\rm L}(r') \frac{r-r'}{r'},
 \end{equation}
 where $H_0$ is the Hubble parameter, $\Omega_m$ is the fractional
matter density parameter, $c$ is the speed of light, $a$ is the scale factor and $n_L(r')$ is probability distribution of the effective number density of galaxies inside a tomographic weak lensing bin.
\par Now Fourier transforming $f(\boldsymbol{x_\perp})$, we write,
\begin{equation} \label{eq:eq2}
    \widetilde f(\boldsymbol{k_\perp}) = \int {\rm d} ^2 \boldsymbol{x_\perp} \ f (\boldsymbol{x_\perp}) e ^ {-i \boldsymbol{k_\perp} \cdot \boldsymbol{x_\perp}}. 
\end{equation}
Substituting Eqn.~\ref{eq:eq1} into Eqn.~\ref{eq:eq2} and writing $\mathcal{U}^f(\boldsymbol{x})$ as the inverse Fourier transform of $\widetilde{\mathcal{U}}^f(\boldsymbol{k'})$ implies
\begin{equation}
\begin{aligned}
    \widetilde f (\boldsymbol{k_\perp}) = \int {\rm d} ^2 \boldsymbol{x_\perp} \int {\rm d} r \ \Bigg[ \int \frac{{\rm d} ^3 \boldsymbol{k}'}{(2 \pi)^3} \widetilde {\mathcal{U}}^f (\boldsymbol{k} ')  e ^ {i \boldsymbol{k}' \cdot \boldsymbol{x}} \Bigg] \\ \times Q^f(r) e ^ {-i \boldsymbol{k_\perp} \cdot \boldsymbol{x_\perp}}.
    \end{aligned}
\end{equation}
Using the plane-parallel approximation, $r = \boldsymbol{x_\parallel}$, it follows that
\begin{equation} \label{eq:eq3}
\widetilde f(\boldsymbol{k_\perp}) = \int_0^\infty \frac{{\rm d} k_\parallel}{\pi} \ \widetilde {\mathcal{U}}^f( \boldsymbol k) \widetilde Q^{f*}(k_\parallel),
\end{equation}
where $\widetilde {Q}^{f *}(k_\parallel)$ is the conjugate of the Fourier transform of $Q^f(r)$, that is,
\begin{equation}
\widetilde Q^f(k_\parallel) = \int_0^{r_{\rm max}} {\rm d} r  \ Q^f(r) e ^ {-i k_\parallel r}.
\end{equation}
Now defining the perpendicular power spectrum, $\mathcal{C}^{f_1 f_2}(k_\perp; r_0)$, for projected fields $f_1$ and $f_2$ at co-moving distance, $r_0$, as
\begin{equation}
\langle f_1(\boldsymbol{k_\perp}) f_2(\boldsymbol{k'_\perp})\rangle = (2 \pi)^ 2\delta (\boldsymbol{k_\perp} - \boldsymbol{k'_\perp}) \mathcal{C}^{f_1 f_2}(k_\perp; r_0),
\end{equation}
it follows from Eqn.~\ref{eq:eq3} that the perpendicular power spectrum is 
\begin{equation}
\begin{aligned}
    \mathcal{C}^{f_1 f_2}(k_\perp ; r_0) = \frac{1}{\pi } \int_0^\infty {\rm d} k_\parallel \ \widetilde  Q^{{f_1}*}(k_\parallel) \widetilde Q^{f_2}(k_\parallel) \\ \times P^{f_1 f_2}(k_\parallel, k_\perp),
\end{aligned}
\end{equation}
where the anisotropic power spectrum between the underlying fields, $P^{f_1 f_2}(k_\parallel, k_\perp)$, is given by
\begin{equation}
    P^{f_1 f_2}(k_\parallel, k_\perp) = \mathcal{P}^{f_1}  \mathcal{P}^{f_2} P(k).
\end{equation}
Then relating $\ell$ to $k_\perp$ following $\ell + 1/2 = k_\perp r_0$ implies that the angular power spectrum, $C^{f_1 f_2}(\ell)$, is 
\begin{equation} \label{eq:cl}
\begin{aligned}
    C^{f_1 f_2}(\ell) = \frac{1}{r_0^2} C\Big(k_\perp = \frac{\ell + 1/2}{r_0}; r_0 \Big)
    \\= \frac{1}{\pi r_0^2} \int_0^\infty {\rm d} k_\parallel \ \widetilde  K^{f_1 f_2} (k_\parallel) P^{f_1 f_2}(k_\parallel, k_\perp),
\end{aligned}
\end{equation}
Here we have found it convenient to define a {\it the radial-mode efficiency kernel} as
\begin{equation} \label{eq:efficiency}
\widetilde K^{f_1 f_2} (k_\parallel) = \widetilde Q^{{f_1}*}(k_\parallel) \widetilde Q^{f_2}(k_\parallel).
\end{equation}
Intuitively the radial mode efficiency kernel is a measure of the angular power spectrum's sensitivity to different $k_\parallel$-modes. This will be important in later sections.

\subsection{The Bernardeau-Nishimichi-Taruya (BNT) Basis}
 The lensing kernel, $q(r)$, is typically broad in $r$, but we would expect the cross-covariance between weak lensing and spectroscopic clustering to be larger if the lensing kernels were narrower. Intuitively this is because narrower kernels allow us to probe smaller radial scales. Thus to maximize the effect of the cross-covariance, we apply the Bernardeau-Nishimichi-Taruya (BNT)~\cite{Bernardeau:2013rda} transformation to the lensing kernels. This change of basis maps the original set of tomographic lensing kernels, $\{q(r) \}$, to a new set of kernels, $\{q_{\rm BNT}(r) \}$, which are narrow in $r$.  In this new basis, the kernels become
\begin{equation}
     Q_{\rm BNT}^f(r)=\left\{
  \begin{array}{@{}ll@{}}
    n_{\rm G}(r), & \text{for}\ \ f=G \\
    q_{\rm BNT}(r), & \text{for} \ f=L.
  \end{array}\right.
\end{equation}
The BNT basis is that natural basis to remove sensitivity to poorly modelled baryonic physics and nonlinear structure growth and we refer the reader to~\cite{Taylor:2018snp,Taylor:2020zcg,Taylor:2020imc,Vazsonyi:2021gxl} for more details on the BNT transform and its applications.

\subsection{Windows and Radial Efficiency Kernels} \label{sec:kernels}

\begin{figure}[!hbt]
\includegraphics[width = \linewidth]{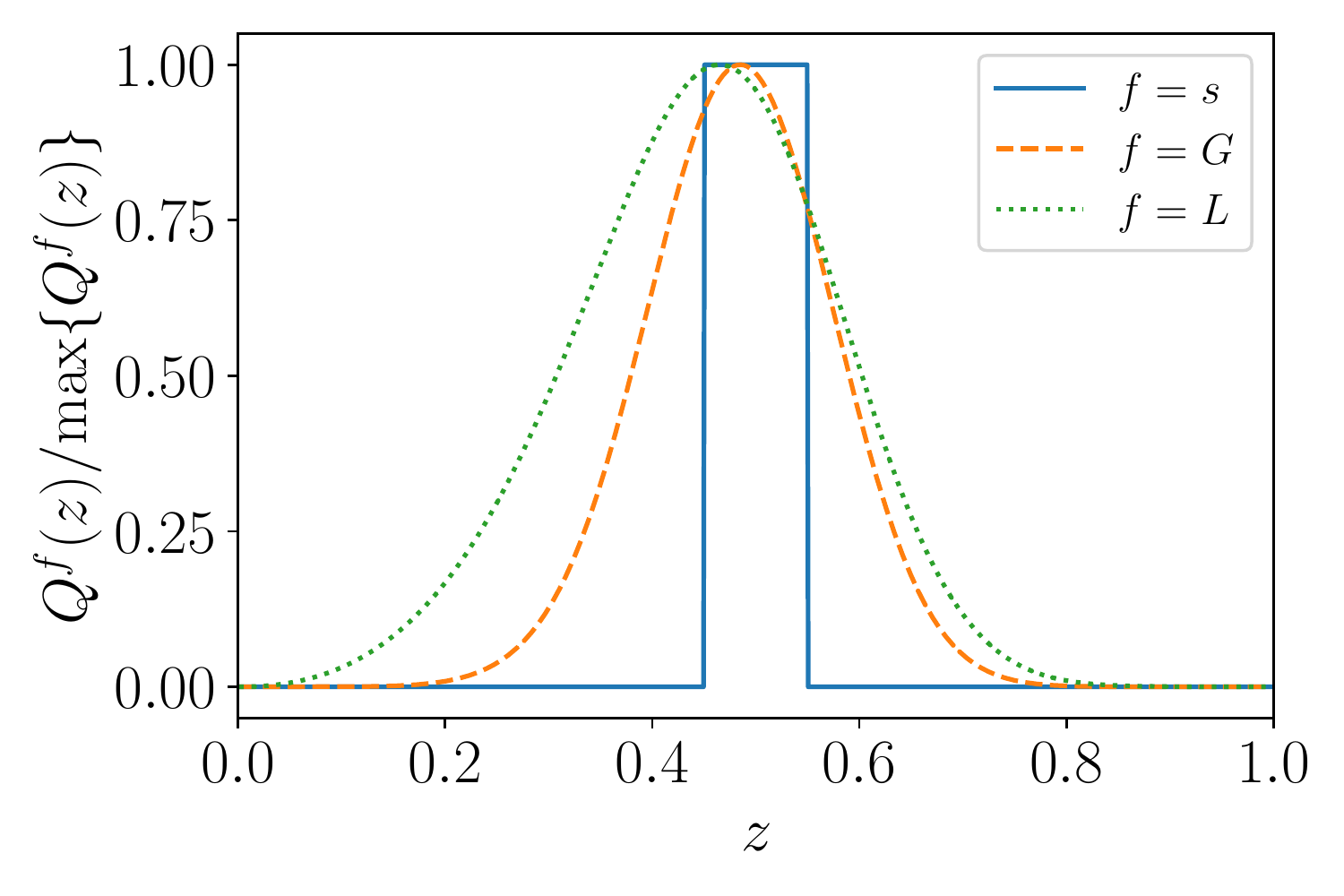}
\caption{The spectroscopic galaxy clustering, photometric galaxy clustering and BNT-transformed weak lensing windows considered in this work. These are labeled by the respective field labels $s$, $G$ and $L$. Since the three fields probe the same cosmological volume we may expect the $3 \times 2$ point statistics estimated from the photometric data to be covariant with the anisotropic power spectrum estimated from the photometric data.}
\label{fig:windows}
\end{figure}

\begin{figure}[!hbt]
\includegraphics[width = \linewidth]{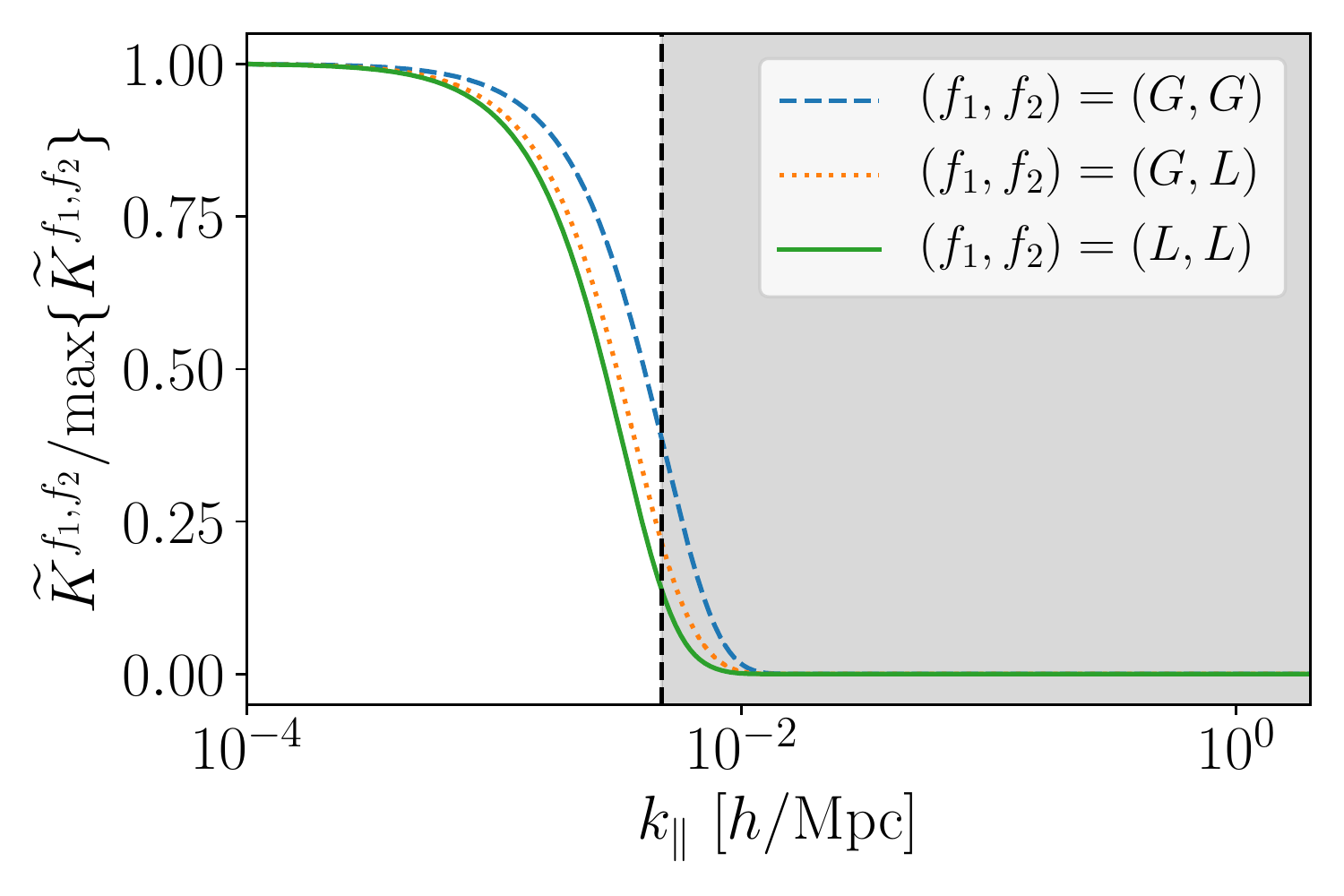}
\caption{Radial-mode efficiency kernels (see Eqn.~\ref{eq:efficiency}) for all three combinations of the photometric clustering and lensing fields. For $P(k_\parallel, k_\perp)$, the largest radial mode (smallest $k_\parallel$) considered in this work is indicated by the dashed line and scales probed by $P(k_\parallel, k_\perp)$ are shaded in gray. For all fields, $\widetilde K^{f_1 f_2} (k_\parallel) \approx 0$ for all $k_\parallel \gtrsim 0.01$ which implies that the photometric $3 \times 2$ point statistics are only sensitive to the largest angular scales. These modes are sample-variance limited, and the overwhelming majority of the information from spectroscopic $P(k_\parallel, k_\perp$) is found at much smaller radial scales suggesting that the covariance between photometric and spectroscopic two-points statistics will have a small impact on paramter constraints. }
\label{fig:kernels}
\end{figure}

Let us consider overlapping spectroscopic, photometric galaxy clustering and weak lensing windows typical of Stage-IV surveys. In this paper we shall assume the same photometric survey window as in~\cite{Taylor:2018snp} which is given by
\begin{equation} \label{eq:survey window}
    n(z) = (z/\bar z) ^2 {\rm exp} \big[ - (z/\bar z) ^{3/2} \big],
\end{equation}
where $\bar z = 0.9$. The galaxies in this window are then equi-partitioned into 10 tomographic redshift bins before being smoothed by a Gaussian kernel with variance $\sigma_z = 0.05 (1+z)$. We calculate the lensing kernels and apply the BNT transform to the resulting windows. In this paper we take the second lowest redshift photometric clustering tomographic bin and the third lowest redshift BNT lensing window. This choice is made so that we probe redshifts near the peak of the lensing kernel where we expect the cross-covariance will be largest.  Meanwhile for the spectroscopic window, we take a top-hat function of width $\Delta z = 0.1$ centred around $z= 0.5$. We take $r_0 = 1959 \ {\rm Mpc}$, corresponding to the co-moving distance at $z=0.5$ throughout the remainder of this work.
\par The resulting windows are shown in Fig.~\ref{fig:windows}. The spectroscopic window lie near the peak of both the photometric galaxy clustering and lensing windows, so that we may na\"ively expect the anisotropic power spectrum estimated from the spectroscopic sample to be strongly covariant with the $3 \times 2$ point statistics estimated from the photometric data.
\par Despite probing the same physical volume the spectroscopic probes and photometric probes are sensitive to very different $k_\parallel$-scales. This can be seen in Fig.~\ref{fig:kernels}, where we plot the radial-mode efficiency kernels, $\widetilde K^{f_1 f_2} (k_\parallel)$, defined in Eqn.~\ref{eq:efficiency} for different field combinations. In particular, we notice that $\widetilde K^{f_1 f_2} (k_\parallel) \approx 0$ for all $k_\parallel \gtrsim 0.01 \ h {\rm Mpc} ^{-1}$ which implies that the $3 \times 2$ point observables are only sensitive to the largest radial scales. Intuitively this is because the photometric redshift error `washes out' sensitivity to small-scale radial modes. 
\par \par Since these large scale modes are sample-variance limited, the overwhelming majority of the information from spectroscopic $P(k_\parallel, k_\perp$) is found at much smaller radial scales~\cite{Chen:2021wdi, Lange:2021zre}. Thus we should qualitatively expect the cross-covariance between the two types of probes to have a small impact on parameter constraints.
\par We will quantify this statement in the following sections and should expect this result for all photometric windows that are broad in $z$. To see why, it is useful to consider a top-hat window with co-moving width $\Delta r$. In this case, the radial-mode efficiency kernel is~\cite{Taylor:2021bhg}
\begin{equation}
    \widetilde K^{f_1 f_2} (k_\parallel) = {\rm sinc}^2 \Big( \frac{k_\parallel \Delta r} {2}\Big),
\end{equation}
where ${\rm sinc(x) = sin(x)/ x}$, so that the maximum $k_\parallel$-scale probed by the window is inversely proportional to the width of the bin, $\Delta r$ (see Fig. 2 in~\cite{Taylor:2021bhg}). This means that the covariance may become important if the photometric tomographic bins are substantially narrower in $\Delta r$ than the bins considered here. We will consider this case in Sect.~\ref{sec:caveats}.

\subsection{2D and 3D Auto-Covariances for Gaussian Fields in the Plane-Parallel Approximation} \label{sec:cov}
In this section we write the analytic expressions for covariances of the photometric and spectroscopic two-point statistics before deriving the cross-covariance between photometric and spectroscopic estimators in the next section. 
\par For Gaussian fields, the covariance of tomographic angular power spectra is found using Wick's Theorem (see e.g the Appendix of~\cite{Zhang:2021wzo}). It is given by
\begin{equation} \label{eqn:cl_cov}
\begin{aligned}
    {\rm Cov} [C^{f_1f_2}(\ell), C^{f_3f_4}(\ell') ] =  \frac{\delta_{\ell \ell'} }{(2 \ell + 1) \Delta \ell f_{\rm sky}} \\ \times \Big( C^{f_1f_3}(\ell)C^{f_2f_4}(\ell) +  C^{f_1f_4}(\ell)C^{f_2f_3}(\ell) \Big)
\end{aligned}
\end{equation}
where the angular power spectra include the shot-noise contribution which is given by
\begin{equation}
     N^{f_1 f_2}(\ell)=\left\{
  \begin{array}{@{}ll@{}}
    \frac{1}{N^{G}}, & \text{for}  f_1 = f_2 = G\\
   \frac{\sigma_\epsilon ^2}{N^L} M_{\rm BNT}^{ij} \delta^{jk} M_{\rm BNT}^{ki} , & \text{for}  f_1 = f_2 = L\\
   0 & \text{otherwise},
  \end{array}\right.
\end{equation}
where $M_{\rm BNT}$ is the BNT transformation matrix (see e.g.~\cite{Taylor:2020zcg} for more details) and we sum over repeated indices, $N^G$ is the effective number of photometric clustering galaxies and $N^L$ is the  effective number of galaxies in the weak lensing sample, and we take the intrinsic ellipticty dispersion, $\sigma_\epsilon = 0.3$ throughout.
\par The covariance of the anistropic power spectrum is also found using Wick's Theorem. It is given by
\begin{equation} \label{eq:pk cov}
\begin{aligned}
    {\rm Cov}[P^{ss}(k_\parallel, k_\perp), P^{ss}(k'_\parallel, k'_\perp)] \\ = \frac{2}{N_k} \left( P^{ss}(k_\parallel, k_\perp) + \frac{1}{N^s}\right) ^2 \delta( \boldsymbol{k} - \boldsymbol{k'} ),
\end{aligned}
\end{equation}
where $N^s$ is the number of spectroscopic galaxies, $N_k$ is the number of modes in the survey volume given by
\begin{equation}
    N_k = 2 k_\perp \Delta k_\perp \Delta k_\parallel \Big( \frac{2 \pi}{V_s ^{1/3}}\Big) ^ {-3},
\end{equation}
where we have ignored the mode coupling from the survey mask and  the volume of the spectroscopic survey, $V_s$, is
\begin{equation} \label{eqn:vs}
    V_s = 4 \pi \int _0 ^ {r_{\rm max}} {\rm d} r \ r^2 n_s(r),
\end{equation}
where the spectroscopic window, $n_s(r)$, is normalized against its maximum value.
\subsection{2D and 3D Cross-Covariance for Gaussian Fields in the Plane-Parallel Approximation}
Now we find an expression for the cross-covariance between the angular power spectra of fields, $f_1$ and $f_2$, and the anistropic power spectrum of the spectroscopic clustering field, $s$.  We define band-powers in $k_\perp$ and $\ell$ such that $\Delta \ell + 1/2 = \Delta k_\perp r_0$ so that photometric and spectroscopic band-powers probe the same perpendicular modes in the plane-parallel approximation. Then using Eqn.~\ref{eq:cl}, which relates angular power spectra to $P(k_\parallel, k_\perp)$, we can write the cross-covariance as
\begin{equation} \label{eq:key1}
\begin{aligned}
   {\rm Cov}[P^{ss}(k_\parallel&, k_\perp), C^{f_1 f_2}(\ell)] = \frac{\delta( k'_\perp - \frac{\ell + 1/2}{r})}{\pi r^2} \\ &\times \int_0^\infty {\rm d} k'_\parallel \ \widetilde  K^{f_1 f_2} (k'_\parallel) \\ &\times {\rm Cov} [P^{s s}(k_\parallel, k_\perp), P^{f_1 f_2}(k'_\parallel, k'_\perp)].
 \end{aligned}
\end{equation}
The problem of finding the covariance is thus reduced to finding ${\rm Cov}[P^{s s}(k_\parallel, k_\perp), P^{f_1 f_2}(k'_\parallel, k'_\perp)]$. This can be found using Wick's Theorem, (see e.g the Appendix of~\cite{Zhang:2021wzo}) 
from which it follows that\footnote{It is useful to notice the similarity with Eqn.~\ref{eq:pk cov}. It is also to note the similarity with which the field indices are paired as in Eqn.~\ref{eqn:cl_cov}, which is a consequence of Wick's Theorem.}
\begin{equation} \label{eq:key2}
\begin{aligned}
    {\rm Cov} [P^{s s}(k_\parallel, k_\perp), P^{f_1 f_2}(k'_\parallel, k'_\perp)] \\= \frac{2}{N^{sf_1f_2}_k}  P^{sf_1}(k_\parallel, k_\perp)P^{sf_2}(k_\parallel, k_\perp) \delta( \boldsymbol{k} - \boldsymbol{k'} ),
\end{aligned}
\end{equation}
where $N^{sf_1f_2}_k$ is the number of modes probed by both the photometric fields, $\{f_1, f_2 \}$, and the spectroscopic field, $s$. For convenience we write $N^{sf_1f_2}_k$ in terms of the number of modes in the spectroscopic window, so that
\begin{equation}
N^{sf_1f_2}_k = \frac{V_{s f_1 f_2}}{V_s} N_k,
\end{equation}
where $V_s$ is the volume of the spectroscopic window given in Eqn.~\ref{eqn:vs} and $V_{s f_1 f_2}$ is the volume probed by all three fields and is given by
\begin{equation}
V_{s f_1 f_2} = 4 \pi \int {\rm d} r  \  r^ 2 {\rm min}_r \{ Q^s(r), Q^{f_1}(r), Q^{f_2}(r)  \}.
\end{equation}
In the above expression, the windows $Q^i(r)$, are normalized against their maximum values. Then from Eqn.~\ref{eq:key1} and Eqn.~\ref{eq:key2} we find that the cross-covariance is
\begin{equation} \label{eq:mixed cov}
\begin{aligned}
    {\rm Cov}[P^{ss}(k_\parallel, k_\perp), C^{f_1 f_2}(\ell)] \\ =  \frac{\delta( k_\perp - \frac{\ell + 1/2}{r_0})}{\pi r_0^2} \frac{  \Delta k_\parallel}{N^{sf_1f_2}_k} \widetilde K^{f_1 f_2} (k_\parallel) \\ \times P^{sf_1}(k_\parallel, k_\perp),P^{sf_2}(k_\parallel, k_\perp) .
    \end{aligned}
\end{equation}
This is the key analytic result of this paper.\footnote{We have carefully accounted for the symmetry of $P(k_\parallel, k_\perp)$ about $k_\parallel = 0$ which introduces a factor of 1/2.}. 
The $\Delta k_\parallel$ appears in the above expression because in practice the integral in Eqn.~\ref{eq:key1} is replaced with a sum when evaluating the power spectrum on a finite grid in radial band-powers, $\Delta k_\parallel$.
\par It is important to notice that cross-covariance is proportional to the radial-mode efficiency kernel $ \widetilde K^{f_1 f_2} (k_\parallel)$. We have already seen that $\widetilde K^{f_1 f_2} (k_\parallel ) \approx 0$ for $k_\parallel \gtrsim 0.01  \ h {\rm Mpc} ^{-1}$ (see Fig.~\ref{fig:kernels}) so that we should only expect the $3 \times 2$  point statistics to be covariant with $P(k_\parallel, k_\perp)$ on the largest radial scales (small $k_\parallel$). 
\par
\begin{figure}[!hbt]
\includegraphics[width = \linewidth]{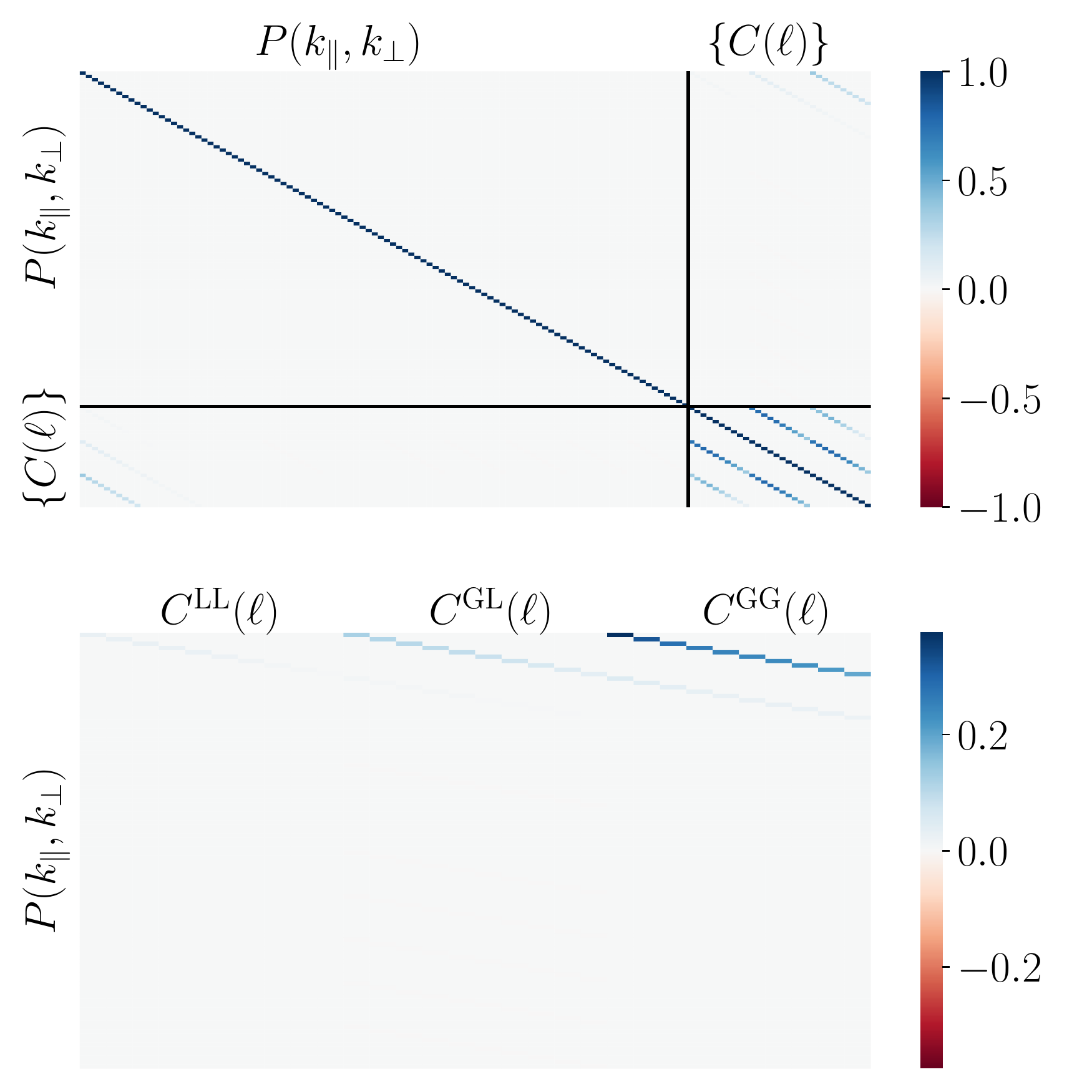}

\caption{{\bf Top:} The correlation matrix of $\{ C(\ell)\}$ and $P(k_\parallel, k_\perp)$. $P(k_\parallel, k_\perp)$ is binned on a $10 \times 10$ log-spaced square grid in  $(k_\parallel,k_\perp)$-space so that $k \in [0.01 \ h {\rm Mpc} ^{-1}, 0.1  \ h {\rm Mpc} ^{-1}]$. The grid in $(k_\parallel,k_\perp)$ is then flattened into the spectroscopic part of the data vector and correlation matrix is arranged so that $k_\parallel$, $k_\perp$ and $\ell$ increase in blocks from left to right and from top to bottom, while $k_\perp$ increases in blocks of fixed $k_\parallel$. {\bf Bottom:} A zoomed-in version of the top panel showing only the cross-correlation between photometric and spectroscopic probes. As expected the covariance is virtually non-existent on all but the largest $k_\parallel$-scales. The information content of these scales is limited due to sample variance i.e, the magnitude of the covariance matrix for low-$k_\parallel$ is large.}
\label{fig:corr}
\end{figure}

 The top panel of Fig.~\ref{fig:corr} shows the correlation matrix between the photometric angular power spectra and the anisotropic power spectrum on a $10 \times 10$ log-spaced square grid in  $(k_\parallel,k_\perp)$-space so that $k \in [0.01 \ h {\rm Mpc} ^{-1}, 0.2  \ h {\rm Mpc} ^{-1}]$ i.e, both $k_\parallel$ and $k_\perp$ lie in the range $[0.007\ h {\rm  Mpc} ^{-1}, 0.14\ h {\rm Mpc} ^{-1}]$. We match the band-powers of the photometric observable to probe the same scales in $k_\perp$ so that $\ell \in [9, 185]$. We use these as our fiducial scale cuts for the remainder of this work. It should be noted that inside this range of scales assuming a Gaussian field to compute the covariance should be an accurate approximation (see e.g~\cite{Barreira:2017fjz} for the photometric case and~\cite{Wadekar:2020hax} for the spectroscopic case).
\par Further we have assumed $10^8$ galaxies per tomographic bin in the photometric case, $10^6$ galaxies in the spectroscopic window\footnote{To avoid double counting galaxies, we further assume that no galaxies are in both the photometric and spectroscopic samples.} and $f_{\rm sky} = 0.3$ in line with forecasts for Stage-IV surveys. We have summarized our survey setup in Table~\ref{tab:1}.
\par In Fig.~\ref{fig:corr}, the matrix is arranged so that $k_\parallel$ and $k_\perp$ (and hence $\ell$) increase in blocks from left to right and from top to bottom, while $k_\perp$ increases in blocks of fixed $k_\parallel$. The matrix is nearly diagonal as expected due to the statistical independence of different $\boldsymbol{k}$  (and $\ell$) modes in the absence of a mask. The bottom panel shows  a zoom-in of the cross-correlation between the spectroscopic and photometric two-point statistics. We have validated that the covariance is semi-positive definite by confirming that the eigenvalues are positive. Several eigenvalues are negative due to numerical noise, but these are all at least 20 orders of magnitude smaller than the largest eigenvalue.
\par As expected the covariance is virtually non-existent on all but the largest $k_\parallel$-scales where the correlation matrix elements are as large as $\sim 0.45$. We do not expect this to have a large impact because the information content of these scales is small due to sample variance. To quantify the impact of neglecting the cross-covariance, we perform a Fisher analysis described in the next section.

\section{Fisher Analysis} \label{sec:results}

\subsection{Fisher Formalism}
\par Given a set of model parameters, $\{ \theta_\alpha\}$, if we assume the data vector follows a Gaussian likelihood and that it is linear in the model parameters, then a good estimate of the marginal error on $\theta_\alpha$, is
\begin{equation}    
    \sigma(\theta_\alpha) = \sqrt{(F^{-1})_{\alpha \alpha}},
\end{equation}
where $F$ is the Fisher matrix which is given by 
\begin{equation}
    F_{\alpha \beta} =  \boldsymbol{ T_{, \alpha}} \boldsymbol {C^{-1} }  \boldsymbol{ T_{, \beta}^T}.
\end{equation}
Here $,\alpha$ denotes the partial derivative of the theory vector with respect to parameter $\theta_\alpha$.
\par In this analysis we take 
\begin{equation}
    \boldsymbol{T} = \Big( P^{ss}(k_\parallel, k_\perp), C^{\rm LL} (\ell), C^{\rm GL} (\ell), C^{\rm GG} (\ell) \Big),
\end{equation}
and the covariance, $\boldsymbol{C}$, to be
\begin{equation}
\boldsymbol{C} = 
\begin{pmatrix}
{\rm Cov}[P^{ss}, P^{ss}] & {\rm Cov}[P^{ss}, C^{f_1 f_2}]\\
{\rm Cov}[P^{ss}, C^{f_1 f_2}] & {\rm Cov}[C^{f_1 f_2}, C^{f_3 f_4}]\\
\end{pmatrix},
\end{equation}
where the expressions for the submatrices (e.g. ${\rm Cov}[P^{ss}, C^{f_1 f_2}]$) can be found in the preceding section.
If we neglect the covariance between the photometric and spectroscopic probes the covariance matrix becomes
\begin{equation}
\boldsymbol{C_*} = 
\begin{pmatrix}
{\rm Cov}[P^{ss}, P^{ss}] & 0\\
0 & {\rm Cov}[C^{f_1 f_2}, C^{f_3 f_4}]\\
\end{pmatrix}.
\end{equation}
We now ask whether it is a valid approximation to set  $\boldsymbol{C} = \boldsymbol{C_*} $.

\subsection{Fiducial Fisher Analysis Results} \label{sec:fisher}

\begin{table*}[hbt!]
\caption{Analysis setup used for the fiducial Fisher analysis. In Sect.~\ref{sec:caveats} we modify this fiducial setup to  explore situations where one might expect the cross-covariance to have a larger impact.}
\label{table:params}
\begin{center}
\begin{ruledtabular}
\begin{tabular}{ lccccccc }
  Parameter &  Value   \\
  \hline
  \hline
  Fraction of Sky Covered ($f_{\rm sky}$)  &  $0.3$ \\
  Number of Lens Source Galaxies $(N^L)$ & $ 10 ^8$ \\
  Number of Photometric Clustering Galaxies $(N^G)$ & $10 ^8$ \\
  Number of Spectrrocopic Clustering Galaxies $(N^s)$ & $10 ^6$ \\
  $\sigma_\epsilon$ & 0.3 \\
  Survey Windows & See Fig.~\ref{fig:windows} \\
  Redshift Uncertainty $(\sigma_z)$ & $0.05 (1+z)$ \\
  Spectroscopic Scale Cuts $( h {\rm Mpc} ^{-1})$ & $k_{\parallel / \perp}  \in [0.007 , 0.14  ] \implies (k_{\rm min}, k_{\rm max}) = (0.01, 0.2)$\\
  Photometric Scale Cuts &  $\ell \in [9, 185]$
\end{tabular}
\label{tab:1}
\end{ruledtabular}
\end{center}
\end{table*}

\begin{figure*}[!hbt]
\includegraphics[width = 0.75 \linewidth]{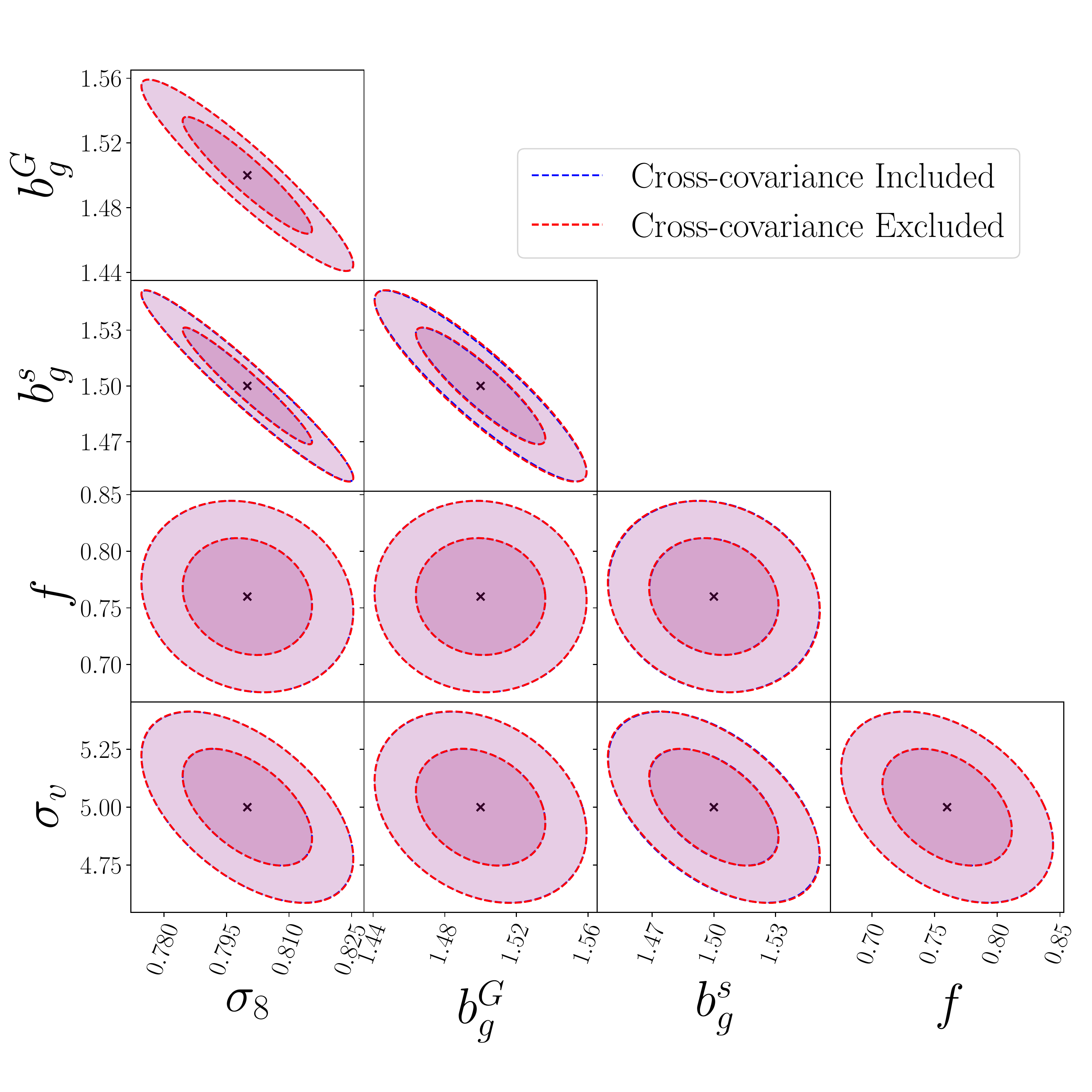}
\caption{The $68 \%$ and $95 \%$ confidence region for the fiducial Fisher analysis (see Sect.~\ref{sec:fisher}). Blue contours are generated using the full cross-covariance between spectroscopic and photometric probes while the red contours are generated after setting the cross-covariance to zero. The two forecasts are in almost exact agreement indicating that the cross-covariance has a negligible effect on parameter constraints.}
\label{fig:triangle}
\end{figure*}

Taking our fiducial analysis setup summarized in Table~\ref{tab:1}, we perform two Fisher analyses constraining the parameters $(f, \sigma_8, b_g^G, b_g^s, \sigma_v)$. In the first analysis we account for the full covariance between the photometric and spectroscopic parts of the data vector, while in the second we take $\boldsymbol{C} = \boldsymbol{C_*} $. 

\par The resulting parameter forecasts are shown in Fig.~\ref{fig:triangle}. The red contours include the contributions from the cross-covariance between the photometric and spectroscopic probes, while the blue contours do not. We find that excluding the cross-covariance results in a less than $1 \%$ change in the measurement error for all parameters. Again this is because even though there is covariance on the largest radial scales (see Fig.~\ref{fig:corr}), in the case of $P(k_\parallel, k_\perp)$, the information content of these modes is severely limited by sample variance. In other words the magnitude of the covariance matrix for low-$k_\parallel$ is large.
\par In the following subsections, we consider two cases where cross-covariance could be relevant.

\subsection{Impact of Scale Cuts and Photometric Uncertainty } \label{sec:caveats}
\par In the fiducial example, the covariance of the photometric and spectroscopic two-point observables is negligible because the two probes are sensitive to different $k_\parallel$-scales (see Fig.~\ref{fig:kernels}). But there are three circumstances in which this covariance may become important. 
 \par The first instance is the case where we choose to impose more conservative $k$-cuts on the anisotropic power spectrum, $P(k_\parallel, k_\perp)$. We may wish to do this when testing a theory of modified gravity where the power spectrum is not known deep into the nonlinear regime. After imposing more conservative cuts, the covariant scales provide a larger percentage of the signal and hence any covariance on these scales may be relevant. We test whether this is the case by repeating the Fisher analysis of the prior section, this time choosing the conservative scale cuts $k \in [0.01 \ h {\rm Mpc} ^{-1}, 0.05  \ h {\rm Mpc} ^{-1}]$. Again we find that neglecting the cross-covariance results in a less than $1 \%$ change on the forecasted parameter measurement errors. 
 \par The second instance, is the case where we probe $k_\perp$ deeper into the nonlinear regime than $k_\parallel$. This is a realistic scenario as nonlinear redshift-space distortion modelling uncertainty means we must take a more conservative radial scale cut compared to the perpendicular direction. To determine whether the cross-covariance is relevant in this scenario we repeat our fiducial Fisher analysis, but this time extend the range of perpendicular scales to include $k_\perp \in [0.007\ h {\rm  Mpc} ^{-1}, 0.5\ h {\rm Mpc} ^{-1}]$, so that $\ell \in [9,656]$. As before, we keep $k_\parallel \in [0.007\ h {\rm  Mpc} ^{-1}, 0.14\ h {\rm Mpc} ^{-1}]$. We find that neglecting the cross-covariance still results in a less than $1 \%$ change in the measurement error of all parameters.
 \par Finally, the cross-covariance may become more important if the photometric tomographic bins in an analysis are narrower than those in our fiducial analysis. This is because the maximum $k_\parallel$-scale probed by the angular power spectra are to a good approximation inversely proportional to the width of the photometric tomographic bin, $\Delta r$ (see Sect.~\ref{sec:kernels}). To test whether this is the case, we repeat the fiducial analysis, choosing the same photometric survey window as in Eqn.~\ref{eq:survey window} but this time we divide the survey window into 20 tomographic bins in the range $z \in [0., 2.5]$ and set the smoothing parameter $\sigma_z = 0$, i.e, an extreme example where there is no photometric redshift uncertainty. The bin width of the resulting tomographic windows is $\Delta z = 0.125$.  Repeating the fiducial analysis with these windows, we find that neglecting the cross-covariance still results in a less than $4 \%$ change in the forecasted errors of all parameters.

\section{Conclusion}
Stage-IV photometric and spectroscopic surveys will probe the same underlying cosmological volume. Hence in the future, we may need to account for the cross-covariance between the two probes when performing cosmological parameter inference. In light of this, we have derived an expression for the covariance between the $3 \times 2$ point photometric angular power spectra and the spectroscopic anisotropic power spectrum under the plane-parallel approximation in the Gaussian field limit. 
\par Assuming a Gaussian Fingers-of-God model, we have found that the two signals are covariant on large radial scales with correlation matrix elements as large as $\sim 0.45$. However the information content of these scales is small due to sample variance so that one may expect the cross-covariance to have a negligible impact on parameter constraints. By performing two Fisher analyses, we have found that the impact of the cross-covariance is indeed negligible and results in a less than $1 \%$ change in the measurement error f cosmological parameters. We have also confirmed that this result is robust to the choice of scale cut and substantially narrower tomographic bins than anticipated for Stage-IV surveys. 
\par Our results extend to different choices of estimators. We have applied the BNT transformation to the lensing kernels, but the standard lensing kernels are even broader and hence even less sensitive to small $k_\parallel$-scales. It follows that neglecting the cross-covariance in the standard analysis has an even smaller impact on parameter constraints. We do not consider other derived estimators like the tomographic angular correlation function, $\xi^{ij} (\theta)$, in the photometric case and Legendre multipoles in the spectroscopic case. However we expect these estimators to be sensitive to the same $k_\parallel$-scales as the underlying anistropic power spectrum, or angular power spectra. Therefor, we can neglect the covariance between any combination of photometric and spectroscopic two-point statistics. 
\par We stress that the results presented in this work are limited to the parameters that affect large scale structure growth in the late Universe and may not apply to measurements of the non-local primordial non-Gaussianity parameter, $f_{\rm NL}$, which is sensitive to the largest scales (see e.g.~\cite{Dore:2014cca}). We have also not tested the impact of the cross-covariance when extracting information from the BAO feature. These extensions are left to a future work.

\section{Acknowledgements}
 The authors thank Alkistis Pourtsidou and Eric Huff for useful discussions, and Anurag Deshpande for providing the plotting routine used to generate the Fisher contours. PLT acknowledges support for this work from a NASA Postdoctoral Program Fellowship. 
 %The authors were supported in part by NASA ROSES 21-ATP21-0050. 
 This research was carried out at the Jet Propulsion Laboratory, California Institute of Technology, under a contract with the National Aeronautics and Space Administration. We acknowledge use of the open source software~\cite{2020SciPy-NMeth, 2020NumPy-Array, 4160265}.

\bibliographystyle{apsrev4-1.bst}
\bibliography{bibtex.bib}

\end{document}